%% file: manuscript.tex
\documentclass[manuscript]{aastex}

\usepackage{graphicx}
\usepackage{natbib}
\usepackage{bm}
\usepackage{wasysym}
\usepackage{verbatim}
\usepackage{rotating}
\usepackage{epsfig}
\usepackage{graphicx}
\usepackage{subfigure}
\usepackage{multirow}

\usepackage{natbib}
\newcommand{\nc}{\newcommand}
\newcommand{\logepso}{\log \epsilon_O}
\newcommand{\logepsni}{\log \epsilon_{Ni}}
\newcommand{\logepsc}{\log \epsilon_{C}}
\newcommand{\teff}{$T_{eff}$}

\newcommand{\nd}{\nodata}
\newcommand{\vsini}{$v \sin i$}
\newcommand{\um}{$\mu$m}
\newcommand{\chiexC}{8.537} 
\nc{\fTellFailO}{53}
\nc{\fTellFailC}{43}
\nc{\errFloor}{0.03}

\newcommand{\nThreeD}{1025}
\newcommand{\nSampStars}{1070} 
\newcommand{\nSpecTot}{15,000}

\input {texcmd.tex}
\newcommand{\waspO}{$0.29_{-0.10}^{+0.06}$} 
\newcommand{\waspC}{$0.10_{-0.06}^{+0.04}$}
\newcommand{\waspCO}{$0.40_{-0.07}^{+0.11}$}
\newcommand{\waspVmag}{11.69}
\newcommand{\waspNobsO}{9}
\newcommand{\waspNobsC}{7}

\newcommand{\colorone}{black}
\newcommand{\colortwo}{blue}
\newcommand{\colorthree}{red}

\begin{document}

\shortauthors{E. Petigura \& G. Marcy}
\shorttitle{Carbon and Oxygen in the Stars}

\title{
Carbon and Oxygen in Nearby Stars:
Keys to Protoplanetary Disk Chemistry\altaffilmark{0}
}

\author{Erik~A.~Petigura}
\affil{Astronomy Department, University of California,
    Berkeley, CA 94720}
\email{epetigura@berkeley.edu}

\author{Geoffrey~W.~Marcy}
\affil{Astronomy Department, University of California,
    Berkeley, CA 94720}

\altaffiltext{0}{Based in part on observations obtained at the
  W.~M.~Keck Observatory, which is operated as a scientific
  partnership among the California Institute of Technology, the
  University of California, and the National Aeronautics and Space
  Administration. The Observatory was made possible by the generous
  financial support of the W.~M.~Keck Foundation.}


\begin{abstract}
  We present carbon and oxygen abundances for \nStarsTot~FGK
  stars---the largest such catalog to date.  We find that
  planet-bearing systems are enriched in these elements.  We
  self-consistently measure $N_{C}/ N_{O}$, which is thought to play a key role in planet formation.  We identify~\ncoGtThresh~stars with $N_{C}/ N_{O}$ $\geq$ \coThresh~as potential hosts of carbon-dominated exoplanets. We measure a downward trend in [O/Fe] versus [Fe/H] and find
  distinct trends in the thin and thick disk, supporting the work
  of~\cite{Bensby04}.
  Finally, we measure sub-solar $N_{C}/ N_{O}$ = \waspCO~for WASP-12, a
  surprising result as this star is host to a transiting hot Jupiter
  whose dayside atmosphere was recently reported to have $N_{C}/ N_{O}$ $\geq$ 1
  by \cite{Madhu11}.
  Our measurements are based on \nSpecTot~high signal-to-noise spectra
  taken with the Keck 1 telescope as part of the California Planet
  Search.  We derive abundances from the [OI] and CI absorption lines
  at $\lambda = $ 6300 and 6587~\AA~using the {\tt SME} spectral
  synthesizer.
\end{abstract}

\section{Introduction}
After primordial hydrogen and helium, carbon and oxygen are the most
abundant elements in the cosmos.  Life on earth is built upon the
versatility of carbon's four valence electrons and is powered by
metabolizing nutrients with oxygen.

The prevalence of carbon and oxygen gives them a prominent role in stellar interiors, opacities, and energy generation.  As a result, studying their abundances helps to reveal the nucleosynthetic chemical evolution of galaxies.  

The interstellar medium is thought to be enriched with oxygen by Type
II supernovae.  Taken with iron, which is produced in both Type Ia and
Type II supernovae, oxygen provides a record of galactic chemical
enrichment and star formation rate~\citep{Bensby04}.  It is well known
that stars synthesize helium into carbon through the triple alpha
reaction.  However, it is still unclear which stars dominate carbon
production in the galaxy.  For a discussion of the possible sites of
carbon synthesis see ~\cite{Gustafsson99}.

The ratio of carbon to oxygen ($N_{C}/ N_{O}$) is thought to play a critical role
in the bulk properties of terrestrial extrasolar
planets. ~\cite{Kuchner05} and~\cite{Bond10} predict that above a
threshold ratio of $N_{C}/ N_{O}$ near unity, planets transition from silicate-
to carbide-dominated compositions.

We present the oxygen and carbon abundances derived from the [OI] line
at 6300~\AA~and the CI line at 6587~\AA~for \nStarsTot~stars in the
California Planet Search (CPS) catalog.  We compute the abundances
with the {\tt Spectroscopy Made Easy (SME)} spectral
synthesizer~\citep{SME}.  Using {\tt SME}, we self-consistently
account for the NiI contamination in [OI] and report detailed Monte
Carlo-based errors.  Others have measured stellar carbon and oxygen
before. \cite{Edvardsson93} measured oxygen in 189 F and G dwarfs,
and~\cite{Gustafsson99} measured carbon in 80 of these stars.  More
recent studies include,~\cite{Bensby05},~\cite{Luck06},
and~\cite{Ramirez07}.  However, the shear number (\nSpecTot) of CPS
spectra give us a unique opportunity to measure the distributions of
these important elements in a large sample.

\section{Observations}
\subsection{Stellar Sample}
The stellar sample is drawn from the Spectroscopic Properties of Cool
Stars (SPOCS) catalog \citep[hereafter VF05]{SPOCS} and from the N2K
(``Next 2000'') sample~\citep{N2K}.  We include 533 N2K stars and 537
VF05 stars for a total of~\nSampStars~stars.

We adopt stellar atmospheric parameters for each star from VF05 and
from the identical analysis for the N2K targets (D. Fischer 2008,
private communication).  These parameters are: effective temperature,
$T_{eff}$; gravity, $\log g$; metallicity, [M/H]; rotational
broadening, $v \sin i$; macroturbulent broadening, $v_{mac}$;
microturbulent broadening, $v_{mic}$; and abundances of Na, Si, Ti,
Fe, and Ni.  Metallically includes all elements heavier than helium.
A star's abundance distribution is the solar abundance pattern
from~\cite{Grevesse98} scaled by the star's metallicity.  Na, Si, Ti,
Fe, and Ni abundances are computed independently from [M/H] and are
allowed vary from scaled solar [M/H].

\subsection{Spectra}
Our spectra were taken with HIRES, the High Resolution Echelle
Spectrograph~\citep{HIRES} between August, 2004 and April, 2010 on the
Keck 1 Telescope. The spectra were originally obtained by the CPS to
detect exoplanets.  For a more complete description of the CPS and its
goals, see~\cite{Marcy08}.  The CPS uses the same detector setup each
observing run and employs the HIRES exposure meter (Kibrick et
al. 2006) to set exposure times, ensuring consistent and high quality
spectra across years of data collection. The spectra have resolution
R~=~50,000 and S/N~$\sim$~200 at 6300 and 6587~\AA.  This analysis
deals with three classes of observations:

\begin{enumerate}
\item {\em Iodine cell in}.  For the majority of its observations, the
  CPS passes starlight through an iodine cell~\citep{Marcy92}, which
  imprints lines between 5000 and 6400~\AA~that serve as a wavelength
  fiducial.  We discuss how we remove these lines and their effect on
  oxygen measurements in \S \ref{sec:iodine}.
\item {\em Iodine cell out}.  Calibration spectra taken without the
  iodine cell.
\item {\em Iodine reference}.  At the beginning and end of each
  observing night, the CPS takes reference spectra of the iodine cell
  using an incandescent lamp.
\end{enumerate}

\section{Spectroscopic Analysis}
\subsection{Line Synthesis}
We use the {\tt SME} suite of routines to fine-tune line lists based
on the solar spectrum, determine global stellar parameters, and
measure carbon and oxygen.  To generate a synthetic spectrum, {\tt
  SME} first constructs a model atmosphere by interpolating between
the~\cite{Kurucz92} grid of model atmospheres.  Then, {\tt SME} solves
the equations of radiative transfer assuming Local Thermodynamic
Equilibrium (LTE).  Finally, {\tt SME} applies line-broadening to
account for photospheric turbulence, stellar rotation, and instrument
profile.  For a more complete description of {\tt SME}, please
consult~\cite{SME} and VF05.  We emphasize that {\tt SME} solves
molecular and ionization equilibrium for a core group (around 400) of
species that includes CO (N. Piskunov 2011, private communication).

\subsection{Atomic Parameters}
Measuring stellar oxygen is notoriously difficult because of the
limited number of indicators in visible wavelengths.  The general
consensus is that the weak, forbidden [OI] transition at 6300~\AA~is
the best indicator because it is less sensitive to departures from
local thermodynamic equilibrium than other indicators.  In dwarf stars, this line suffers from a significant NiI blend, which is 
an isotopic splitting of $^{58}$Ni and $^{60}$Ni \citep{Johansson03}.  
The NiI feature was first noted by~\cite{Lambert78}, but only recently
included in abundance studies~\citep{Prieto01}.  Carbon is more generous to visual spectroscopists.  We select the CI line at
6587~\AA~because it sits relatively far from neighboring lines and is
in a wavelength region with weak iodine lines (see \textsection~\ref{sec:iodine}).

Line lists are initially drawn from the Vienna Astrophysics Line
Database~\citep{VALD}.  We tune line parameters by fitting the 
disk-integrated National Solar Observatory (NSO) solar spectrum of~\cite{Kurucz84} with the {\tt SME} model of the solar atmosphere.  
Table~\ref{tab:solpar} lists the atmospheric parameters adopted when 
modeling the sun. We fit a broad spectral range from 6295 to 6305~\AA~surrounding the
[OI] line and 6584 to 6591~\AA~surrounding the CI line.  We adopt the solar abundances of~\cite{Grevesse98} except for O and Ni where we adopt $%
\logepso$\footnote{$\log \epsilon_X = \log_{10} (N_X /N_H ) + 12$}
= 8.70 and $\logepsni = 6.17$ ~\citep{Scott09} and $\logepsc = 8.50$
~\citep{Caffau10}.  We adjust line centers, van der Waals broadening
parameters ($\Gamma_6$), and oscillator strengths ($\log gf$) so our
synthetic spectra best match the NSO atlas.  Table~\ref{tab:atomic}
shows the best fit atomic parameters after fitting the NSO solar
atlas.

Given the high quality of the solar spectrum, solar abundances
and line parameters are often measured using sophisticated 
three-dimensional, hydrodynamical, non-LTE codes.  
For this work, however, we are more interested in 
self-consistently determining line parameters using {\tt SME}
than from a more sophisicated solar model.
As a result, the line parameters in Table~\ref{tab:atomic} are not in tight
agreement with the best laboratory measurements.  For example, 
\cite{Johansson03} measured $\log gf = -2.11$ for the NiI blend
in contrast to $\log gf = -1.98$ in this work.  The purpose of fitting the
atomic parameters in the sun is to determine the best parameters given our
atmospheric code and our adopted solar abundance distribution.

\begin{figure}
\begin{center}
\includegraphics{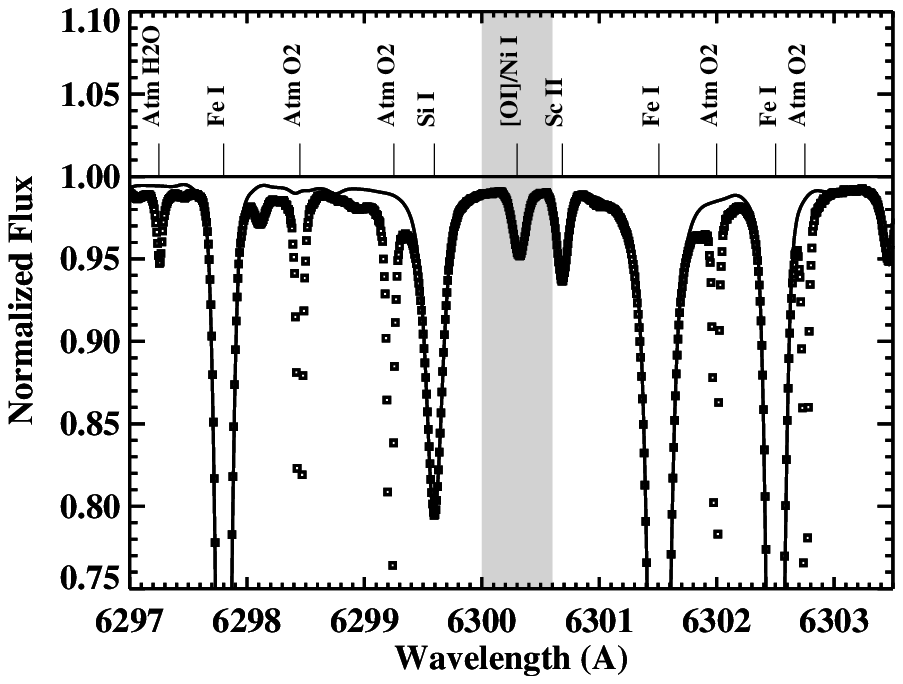}
\end{center}
\caption{Solar spectrum in the vicinity of the [OI] line at 6300.312~\AA.  The points are from the NSO solar atlas, and the solid line is the {\tt SME} fit.  The shaded region marks the region that is included in the $\chi^2$ fit to the [OI]/NiI blend.}
\label{fig:solar_6300}
\end{figure}

\begin{figure}
\begin{center}
\includegraphics{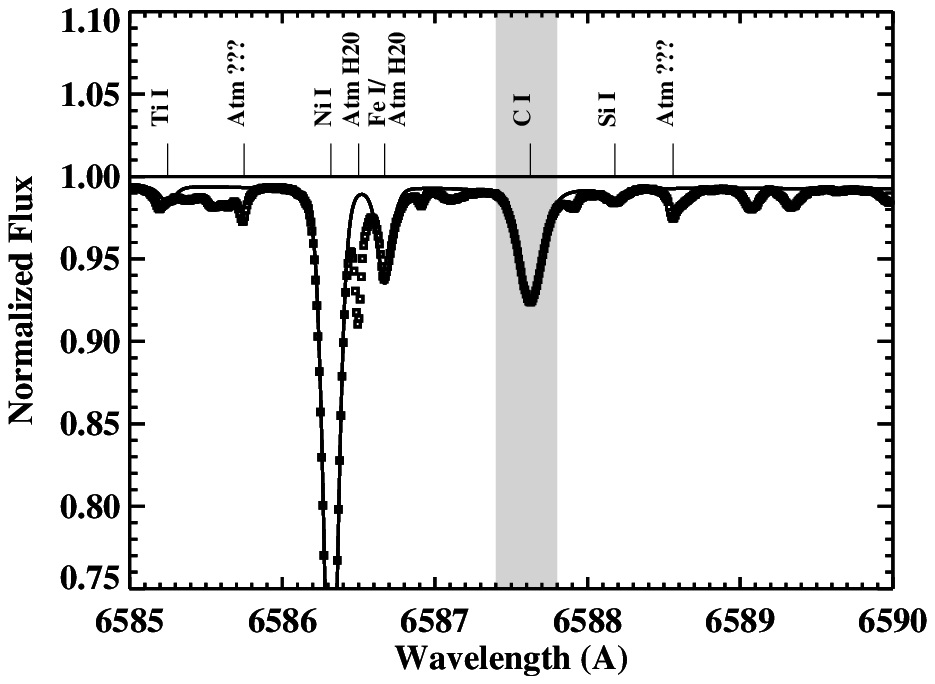}
\end{center}
\caption{Solar spectrum in the vicinity of the CI line at 6587.625~\AA.  The points are from the NSO solar atlas, and the solid line is the {\tt SME} fit.  The shaded region marks the region that is included in the $\chi^2$ fit to the CI line.}
\label{fig:solar_6587}
\end{figure}

We show the fitted NSO spectrum for both wavelength regions in
Figures~\ref{fig:solar_6300} and~\ref{fig:solar_6587}.  The shaded
regions (6300.0-6300.6 and 6587.4-6587.8~\AA) represent the fitting
region.  Only points in the fitting region are used in the $\chi^2$
minimization routines (see \S~\ref{sec:fitting}).

\begin{figure}
\begin{center}
\includegraphics{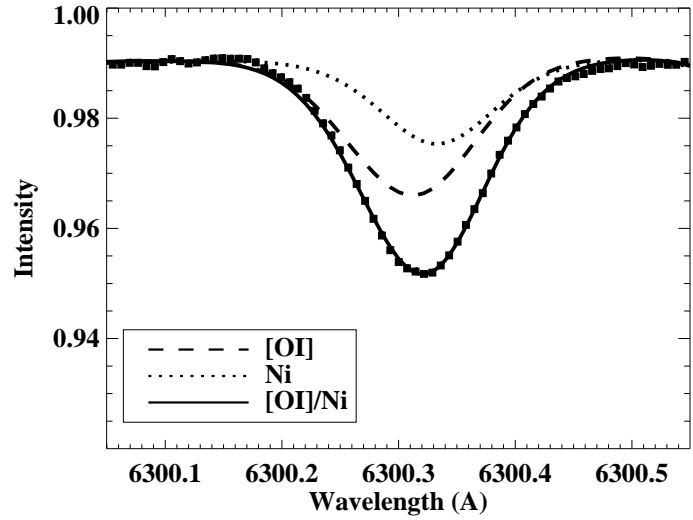}
\end{center}
\caption{The [OI]/NiI blend in the NSO spectrum.  The points are the from NSO solar atlas, and the solid line is the {\tt SME} fit.  The relative contribution of [OI] and Ni are shown by the dashed and dotted lines respectively.}
\label{fig:solarzoom_6300}
\end{figure}

Figure~\ref{fig:solarzoom_6300} shows a close up view of the [OI]/NiI
blend in the sun.  To help the reader visualize the relative contributions
of each line in the sun, we synthesize the oxygen and nickel lines individually.
To compute the relative strength of [OI], we remove all Ni in our solar
model and re-synthesize the spectrum in {\tt SME}.  
To calculate the NiI contribution we remove all oxygen.  
Since the both lines are weak ($<$ 5 \% of continuum), the line profile
for the [OI]/Ni blend is nearly the product of the individual [OI] and Ni lines.
This would not be true in the case of deeper lines.
In the sun, the [OI] and NiI contributions to the blend are comparable.  
In some stars, the blend is decidedly nickel-dominated, while in others,
oxygen dominates.

\begin{figure}
\begin{center}
\includegraphics{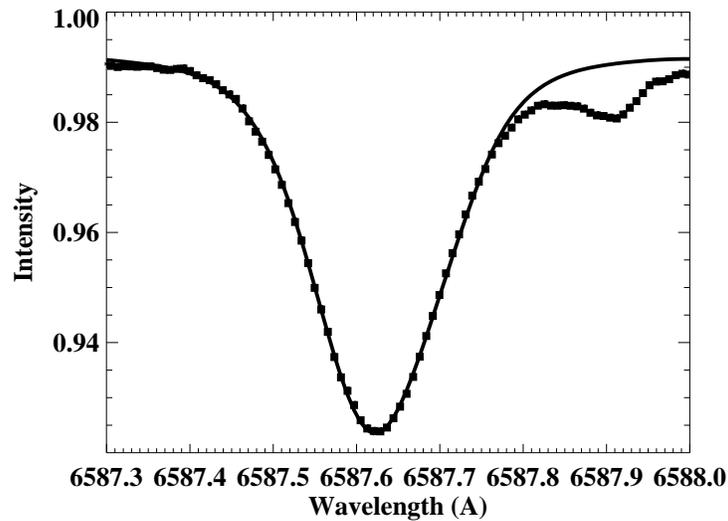}
\end{center}
\caption{The CI line in the NSO spectrum.  The points are the from NSO solar atlas, and the solid line is the {\tt SME} fit.  The fitting region for this line is 6587.4-6587.8~\AA~and excludes the unidentified feature at 6587.9.}
\label{fig:solarzoom_6587}
\end{figure}

Figure~\ref{fig:solarzoom_6587} shows the carbon indicator plotted on the same intensity scale as the oxygen detail shown in Figure~\ref{fig:solarzoom_6300}.  There is an unknown line on the red wing of the carbon indicator.  We exclude the mystery line from the fitting region.  

\begin{figure}
\begin{center}
\includegraphics{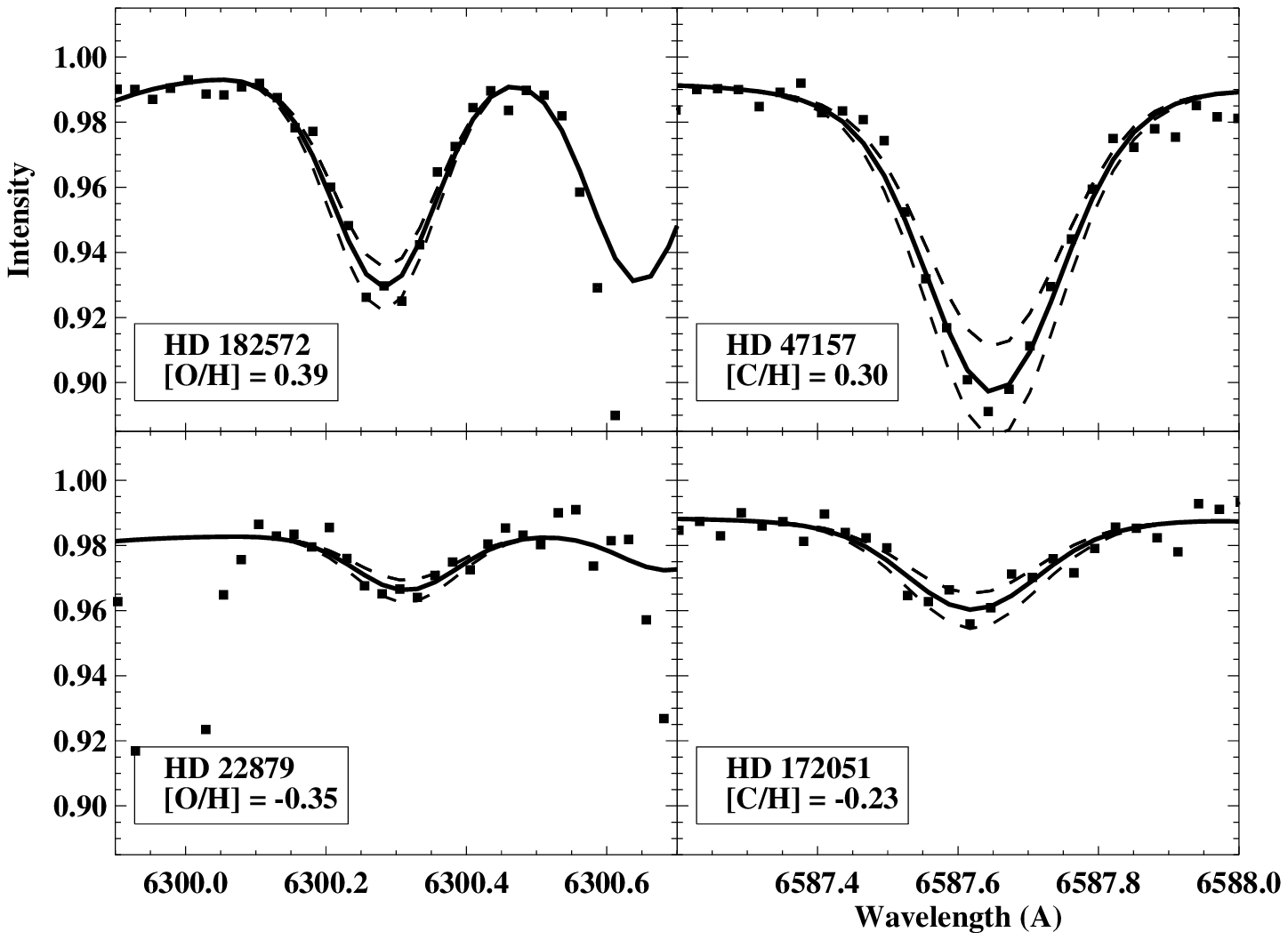}
\end{center}
\caption{Sample spectra of stars with low and high carbon and oxygen abundances.  The solid line shows the best fit {\tt SME} spectrum.  The dashed lines are the {\tt SME} spectra with [X/H] increased and decreased by 0.1 dex $\sim 25$\% from the best fit value.  The [OI] line in the HD~22879 spectrum sits between two telluric lines.
 }
\label{fig:starsamp}
\end{figure}

As a point of reference for the reader, we include stellar counterparts 
to Figures~\ref{fig:solarzoom_6300} and ~\ref{fig:solarzoom_6587} in 
Figure~\ref{fig:starsamp}.  We show stars with low and high carbon and oxygen 
abundance along with the best fit {\tt SME} spectrum.

\begin{deluxetable}{l c }
\tablewidth{0pc}
\tablecaption{Adopted solar atmospheric parameters}
\tablehead{
\colhead{Parameter}& 
\colhead{Value}    \\
}
\startdata
\teff     & 5770 K      \\
$\log g$  & 4.44 (cgs) \\
$v_{mic}$ & 1.00 km/s   \\
$v_{mac}$ & 3.60 km/s   \\
\vsini    & 1.60 km/s   \\
$v_{rad}$ & 0.02 km/s   \\
\enddata
\tablecomments{Adopted atmospheric parameters in the {\tt SME} solar model}
\label{tab:solpar}
\end{deluxetable}

\begin{deluxetable}{l c c c}
\tablewidth{0pc}
\tablecaption{Atomic parameters from fitting the NSO atlas}
\tablehead{
\multirow{2}{*}{Element} &
\colhead{$\lambda$}	&
\multirow{2}{*}{$\log gf$} &
\multirow{2}{*}{$\Gamma_6$}  		   \\
&
\colhead{(\AA) }	&
&
\\
}
\startdata
\sidehead{[OI] region}
\hline
Fe 1 & 6297.801 & -2.766 & -7.89 \\
Si 1 & 6297.889 & -2.899 & -6.88 \\
O  1 & 6300.312 & -9.716 & -8.89 \\
Ni 1 & 6300.335 & -1.983 & -7.12 \\
Sc 2 & 6300.685 & -2.041 & -8.01 \\
Fe 1 & 6301.508 & -0.793 & -7.53 \\
Fe 1 & 6302.501 & -0.972 & -7.99 \\
\sidehead{CI region}
\hline
Ti 1 & 6585.249 & -0.399 & -7.56 \\
Ni 1 & 6586.319 & -2.775 & -7.68 \\
Fe 2 & 6586.672 & -2.247 & -7.76 \\
C  1 & 6587.625 & -1.086 & -7.19 \\
Si 1 & 6588.179 & -3.082 & -7.12 \\
\enddata
\tablecomments{Best fit line center ($\lambda$), oscillator strengths
  ($\log gf$), and van der Waals broadening parameter ($\Gamma_6$) for
  our [OI] and CI indicators and nearby lines.  They are derived by
  fitting the NSO atlas.}
\label{tab:atomic}
\end{deluxetable}

\subsection{Telluric Rejection}
There are several telluric lines from O$_2$ and H$_2$O in the vicinity
of our indicators including the 6300.3~\AA~airglow (see
Figures~\ref{fig:solar_6300} and~\ref{fig:solar_6587}).  These lines
are produced in the rest frame of the Earth and contaminate different
parts of a star's spectrum depending on the relative line of sight
velocity between the Earth and the star.  We compute this velocity
directly from the spectra, by cross-correlating the stellar spectra
with the NSO solar atlas.  Based on this velocity, we account for any
shift in the location of the telluric line in the stellar rest frame.

\begin{figure}
\begin{center}
\includegraphics{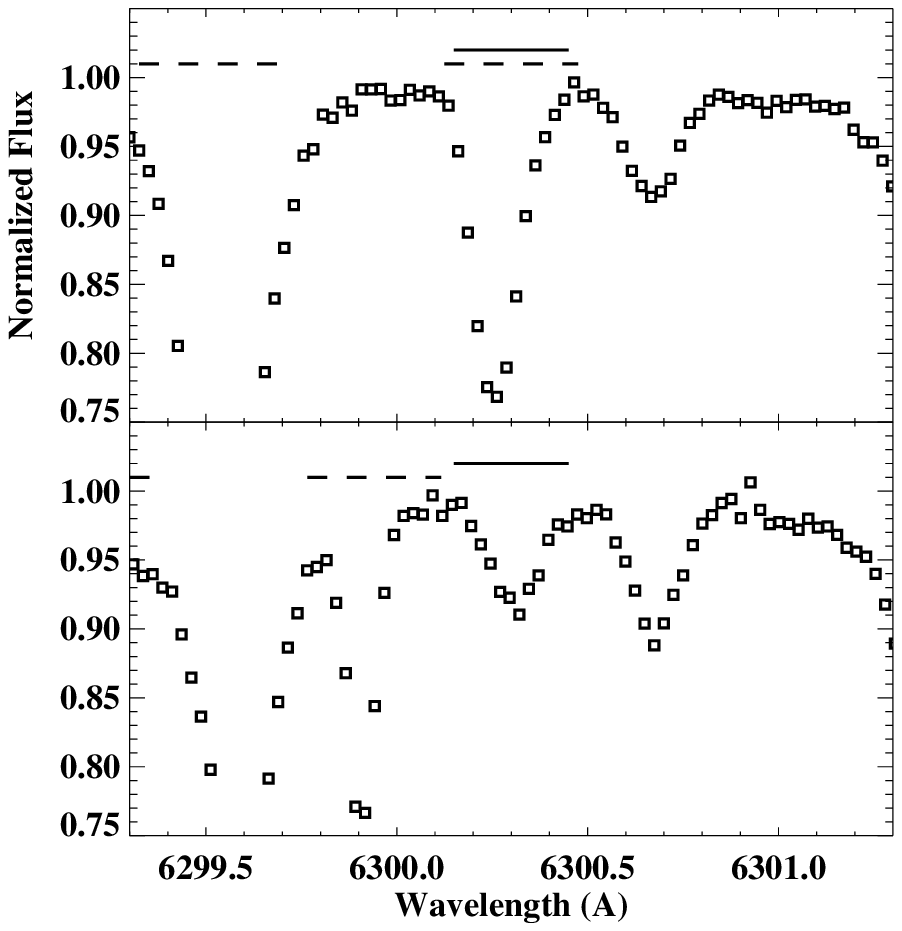}
\end{center}
\caption{Two spectra of HIP~92922 in the star's rest frame.  The [OI]/NiI region (marked with the solid line) in the upper panel is contaminated with an atmospheric O2 (marked with a dashed line).  The same telluric absorption line is shifted 0.3~\AA~blueward in the lower panel and does not contaminate the [OI]/NiI blend.}
\label{fig:tell}
\end{figure}

If a telluric line enters the fitting region, we discard that
observation.  Figure~\ref{fig:tell} shows the [OI]/NiI blend from two
different observations of HIP~92922: one where the blend is
contaminated by a telluric absorption line and one where the blend is
free from telluric contamination.  We reject~\fTellFailO\% of our [OI]
spectra and~\fTellFailC\% of our CI spectra because of telluric
contamination.  Telluric lines affect the [OI] region more strongly
due to the airglow at 6300~\AA.

\subsection{Iodine Removal}
\label{sec:iodine}
\begin{figure}
\begin{center}
\includegraphics{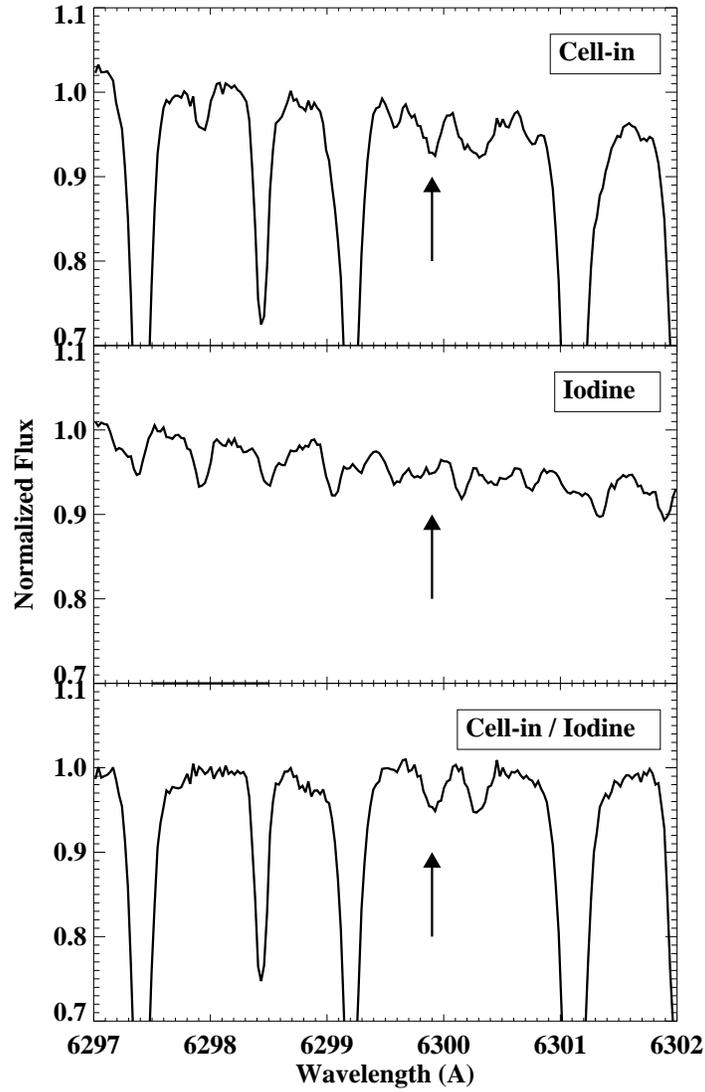}
\end{center}
\caption{HD~148284 spectrum with iodine contamination (top), reference
  iodine observation (middle), and stellar spectrum divided by iodine
  spectrum (bottom).  The arrow marks the [OI]/NiI blend.  The iodine
  is removed to a level of $\sim$1\% of the continuum intensity.}
\label{fig:idiv}
\end{figure}

The majority of the spectra in the CPS catalog were taken through the
iodine cell.  Iodine lines are $\sim 0.5$\% deep in the CI
region---comparable to the photon noise.  In the [OI] region, they are
a $\sim 5$\% effect and must be removed.  For a given iodine cell in
observation, we locate the most recent iodine observation (usually at
the beginning of the night).  We account for any shift of the CCD
between the two observations by cross-correlating spectral orders 8,
9, and 10 ($\lambda =$ 5608 - 5895\AA, where the iodine lines are
strongest).  After removing any shift, we divide the iodine cell in
observations by the iodine reference
observations. Figure~\ref{fig:idiv} shows a stellar spectrum, an
iodine spectrum, and the ratio of the two.

\begin{figure}
\begin{center}
\includegraphics{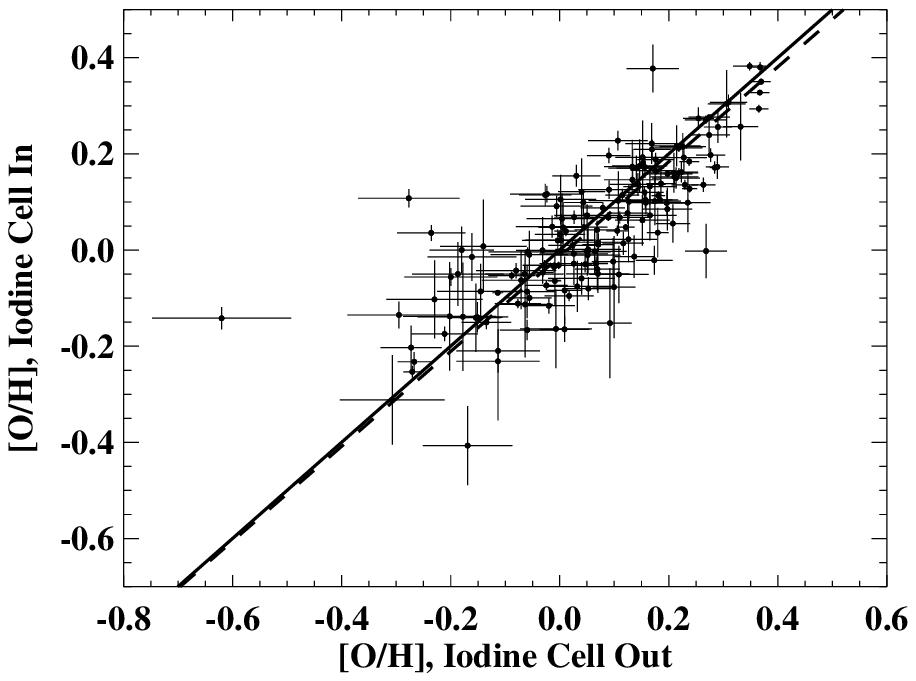}
\end{center}
\caption{Comparison of oxygen abundances derived from spectra taken
  with the iodine cell in (vertical axis) to those taken with the
  iodine cell out (horizontal axis).  The solid line corresponds to an
  equality of the two, while the dashed line shows the best fit to the
  points.  Apparently the oxygen abundances derived from both types of
  spectra show no systematic difference, indicating that the removal
  of the iodine spectrum works without introducing a systematic
  error.}
\label{fig:calib}
\end{figure}

Dividing the iodine spectrum from an iodine cell in spectrum cannot be
done to within photon statistics.  On nights of good seeing, a star's
image may be narrower than the HIRES slit.  The reference iodine
spectra are produced with a lamp that fills the slit uniformly, so the
iodine lines from the iodine cell in observations can be narrower than
the reference iodine lines.  The result is artifacts from the division
at the $\sim1$\% level.  Iodine cell in spectra for a single star
generally yield a larger spread in derived oxygen abundance compared
to iodine cell out observations.  However, when we plot oxygen
abundances derived from iodine cell in observations against abundances
from iodine cell out observations in Figure~\ref{fig:calib}, we see no
systematic trend.

\subsection{Fitting Abundances}
\label{sec:fitting}
We converge on abundance by iterating three $\chi^2$ minimization
routines that fit the continuum, line center, and abundance.  Only
points within the fitting range are used to calculate the $\chi^2$
statistic.  Any point deviating from the fit by more than five times
the photon noise is not included in calculating $\chi^2$.  A short
description of each routine is given below:

\begin{enumerate}
\item {\em Continuum}.  Given the shallowness of our indicators, a
  small error in the continuum level will have a significant effect on
  the derived abundance.  We refine the continuum value by registering
  the level of the spectrum so that $\chi^2$ is minimized.

\item {\em Line center}.  The wavelength zero point and dispersion is
  initially determined from a thorium lamp calibration taken each
  night and refined by cross-correlating the observed spectrum with
  the solar spectrum.  We adjust the radial velocity of the model
  spectrum to minimize $\chi^2$.

\item {\em Abundance}.  We begin with the solar oxygen abundance
  scaled by the star's metallicity.  We refine this value by searching
  over 2 dex of abundance space and minimizing $\chi^2$.
\end{enumerate}

We terminate the iteration when the fits arrive at a stable solution
or when we exceed 10 iterations.

\section{Results}
\subsection{Carbon and Oxygen Abundances}
\label{sec:abundances}
We report [O/H]%
\footnote{[X/H] =  $\log \epsilon_X - \log \epsilon_{X,\odot}$}
and [C/H] for \nStarsO~and \nStarsC~stars respectively.  These are
subsamples of our initial \nSampStars~star sample and arise after we
apply the following global cuts:

\begin{enumerate}
\item {\em \vsini}. In rapidly rotating stars, our indicators can be
  polluted by the wings of neighboring lines due to rotational
  broadening.  When this happens, the abundances of our elements of
  interest become degenerate with that of the polluting line.  This
  effect sets in earlier for the [OI] line, which sits shoulder to
  shoulder between SiI and ScII features. We do not report oxygen or
  carbon abundances for stars with \vsini~greater than \vsiniCutO~and
  \vsiniCutC~km/s respectively.
\item {\em \teff}.  The high excitation energy of the CI line
  (\chiexC~eV) means the line is very weak in cool stars.  For
  example, at 5000K, the line depth in a solar analog is 1\%.  We do
  not report carbon abundances for stars cooler than 5300 K.
\item {\em Statistical scatter}.  We choose to report abundances for
  stars where the scatter in derived abundance is less than
  \scatterCut~dex or, in other words, stars where our measurements are
  precise to a factor of 2.  Our estimates of measurement precision
  are based on empirical scatter and a Monte Carlo analysis, which we
  describe in sections \S~\ref{sec:staterr} \&~\S~\ref{sec:nierr}.
  While our measurement precision is based on a variety of factors
  including line depth and signal to noise, stars that fail this cut
  generally have sub-solar carbon and oxygen abundances.
\end{enumerate}

\begin{figure}
\begin{center}
\includegraphics{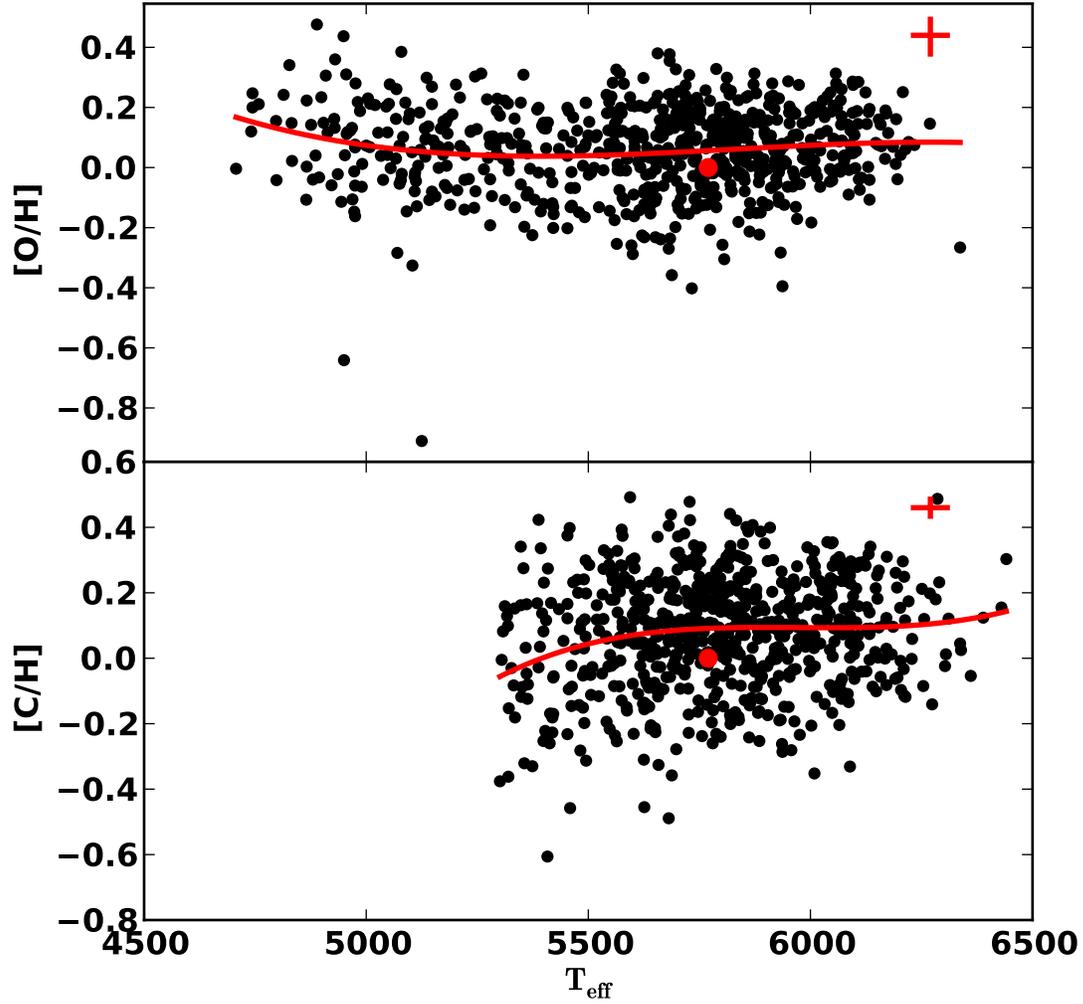}
\end{center}
\caption{Plots showing systematic trends of [O/H] and [C/H] with
  temperature.  The red line is the best fit cubic.  Our correction for
  the \teff trend is this cubic with the constant term chosen so that
  the correction is zero at $T_{eff} =$ \teffSol~K (large red dot).  The
  crosses show the median errors.}
\label{fig:teff}
\end{figure}

With our large stellar sample, it is possible to detect and correct
for systematic trends that would be invisible in smaller samples.
Figure~\ref{fig:teff} shows carbon and oxygen abundances plotted
against temperature.  We believe that the~\cite{Kurucz92} model atmospheres
are most accurate for solar analogs and that errors in the atmosphere
profile grows as we move away from $T_{eff} = T_{\odot}$ = ~\teffSol~K.

We model the systematic behavior of implied abundance
with \teff~by fitting a cubic to the data.  Simply subtracting
out the cubic would artificially force the mean [X/H] to zero, but there
is no reason why the mean disk abundance should be solar. 
Therefore, we let the solar abundance fix the constant term in the cubic by
requiring the systematic correction be zero at~\teffSol~K.  
This correction reaches \maxTO~dex for oxygen and
\maxTC~dex for carbon.  We have removed the temperature trend for all
abundances quoted henceforth.

By removing abundance trends with \teff~for the sake of correcting errors in
atmosphere models, we may
have erased a real astrophysical trend of [X/H] with~\teff.  For example,
hotter stars are more massive and have shorter main sequence lifetimes than
cool stars.  Therefore, the hotter stars in our sample are on average younger
and formed at a later time in the galactic chemical enrichment history.
However, we chose to remove the~\teff~trends because we believe 
uncertainties in atmospheric models are the dominant effect. 

\begin{figure}
\begin{center}
\includegraphics{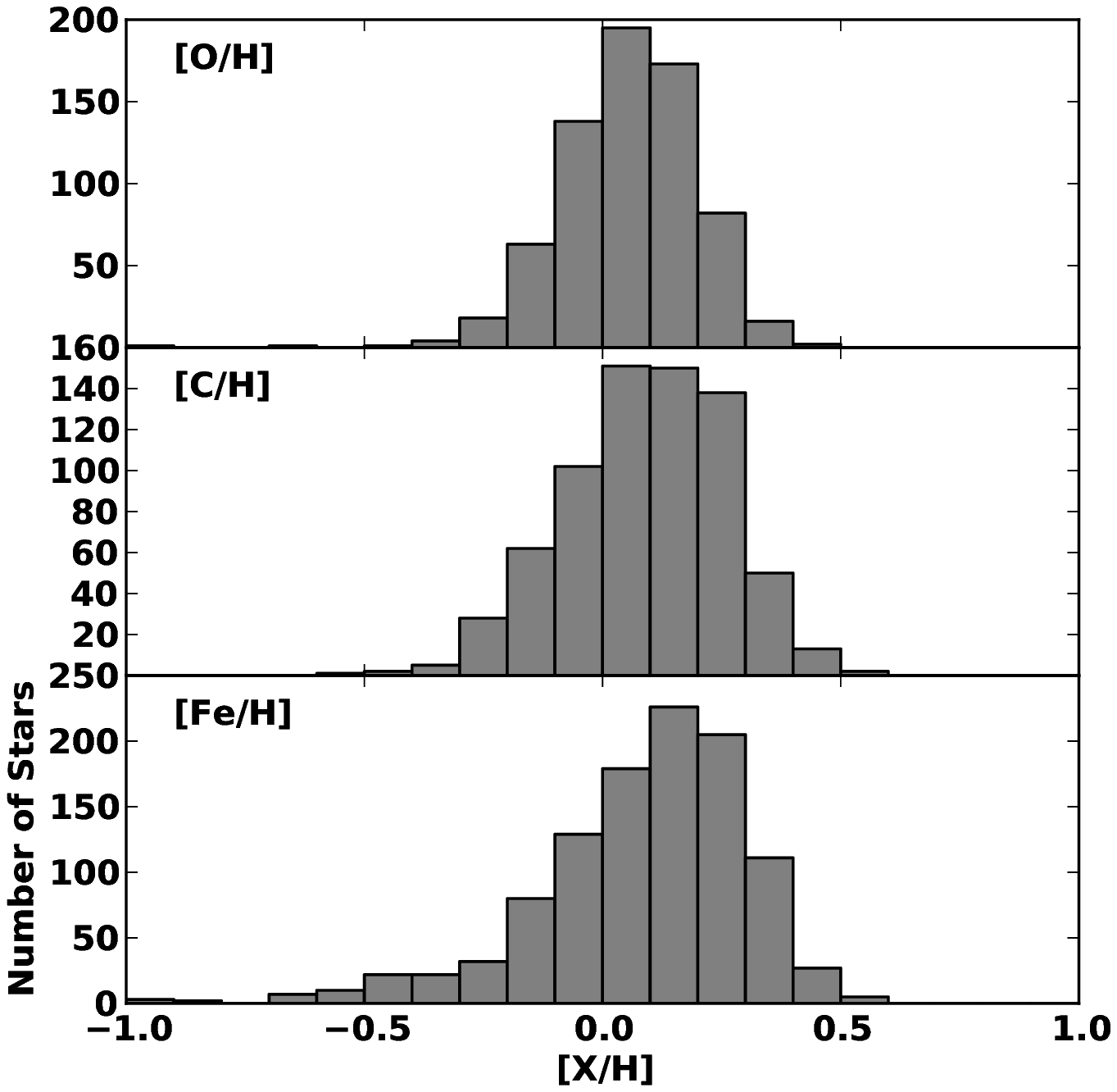}
\end{center}
\caption{Distributions of [O/H], [C/H], and [Fe/H] for comparison.}
\label{fig:abundhist}
\end{figure}

We report our temperature-corrected values for [O/H] and [C/H] with
85\% and 15\% confidence limits along with other stellar data in the
Appendix.  We summarize the statistical properties of derived
abundances in Table~\ref{tab:abundhist} and show their distributions
in Figure~\ref{fig:abundhist}.

\begin{deluxetable}{l c c c c c}
\tablewidth{0pc}
\tablecaption{Summary of Derived Abundances.}
\tablehead{
\colhead{}      &
 \multirow{2}{*}{N} &
\colhead{m}	&
\colhead{S}	&
\colhead{Min}	&
\colhead{Max} \\
\colhead{}      &
\colhead{}	&
\colhead{(dex)}	&
\colhead{(dex)}	&
\colhead{(dex)}	&
\colhead{(dex)} \\
}
\startdata
\input{abundhist.tex}
\enddata
\tablecomments{Here, N is the number of stars with determined
  abundances, m is the mean abundance, and S is the standard deviation
  of abundance distribution.}
\label{tab:abundhist}
\end{deluxetable}

\subsection{Random Errors}
\label{sec:staterr}
We use Monte Carlo bootstrapping to estimate random errors.  We
generate Monte Carlo spectra by scrambling the residuals from our fits
and adding them back to the synthetic spectra.  For each star we
generate and refit 1000 Monte Carlo realizations of the spectrum.  The
resulting abundance distribution provides a good estimate of the true
error distribution.

\begin{figure}
\begin{center}
\includegraphics{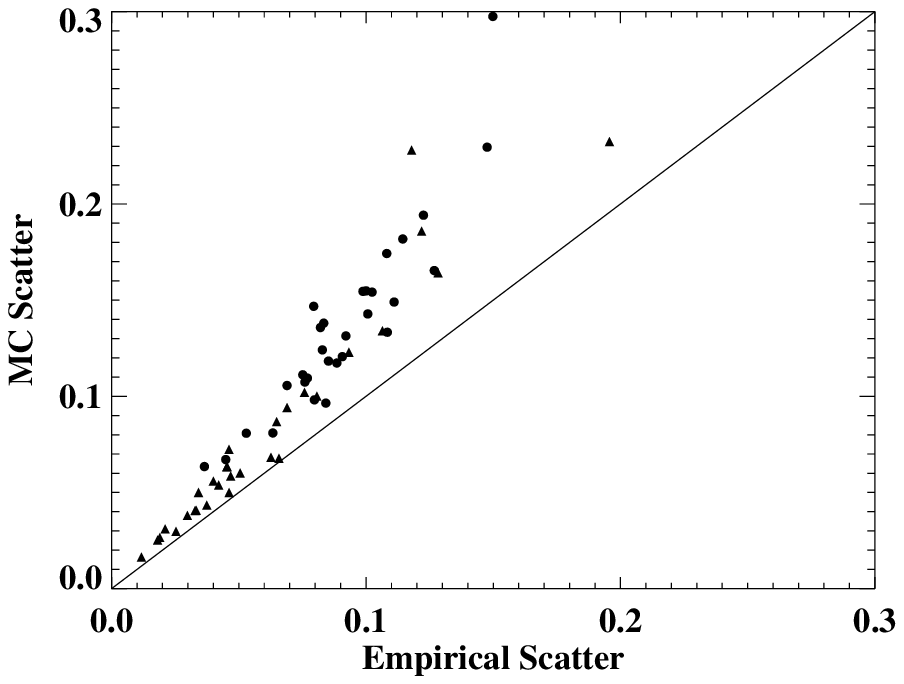}
\end{center}
\caption{Scatter in MC simulations as a function of empirical scatter
  for stars with more than 50 observations.  Oxygen and carbon
  measurements are represented by circles and triangles respectively.
  The solid line represents a 1:1 correlation.}
\label{fig:mcstaterr}
\end{figure}

For some stars we have many independent spectra, allowing us to
compute confidence limits of the oxygen abundances from them as an
{\em empirical} measure of our internal errors.
Figure~\ref{fig:mcstaterr} shows the length of the error bars computed
empirically and from Monte Carlo for stars with more than 50 empirical
fits.  The error estimate from Monte Carlo tracks the empirical
scatter well, slightly overestimating it.  This is due to systematic
errors in our fits that appear as random errors when we scramble the
residuals.

For stars with fewer than 20 observations, we adopt the Monte Carlo confidence intervals as our statistical error; for stars with 20 or more observations, we adopt the empirical confidence intervals.  We diminish these errors by $\sqrt{N_{obs}}$.  Futhuremore, we impose an error floor of \errFloor~dex.

\subsection{Nickel Systematics}
\label{sec:nierr}
Since we are deriving oxygen from a line that is blended with nickel,
the errors in nickel abundance are covariant with errors in oxygen
abundance.  FV05 quote a uniform error of 0.03 dex for their nickel
measurements.  The amount that [OI] and NiI contribute to the blend is
different for every star.  Therefore, we evaluate the effect of the
0.03 dex error in nickel abundance on oxygen abundance on a
star-by-star basis.  We begin with a synthetic spectrum at our quoted
oxygen abundance.  We then refit the oxygen line to a spectrum with
0.03 dex more and 0.03 dex less nickel.  These errors are added in
quadrature to the statistical errors.

There are many other sources of systematic error in our abundance
measurements such as inaccurate solar reference abundances, additional
blends, and our assumption of LTE.  However, these effects should be
largely consistent between stars, so we expect them to contribute
little to errors in our differential abundances.

\subsection{Comparison with Literature}
\begin{figure}
\begin{center}
\includegraphics{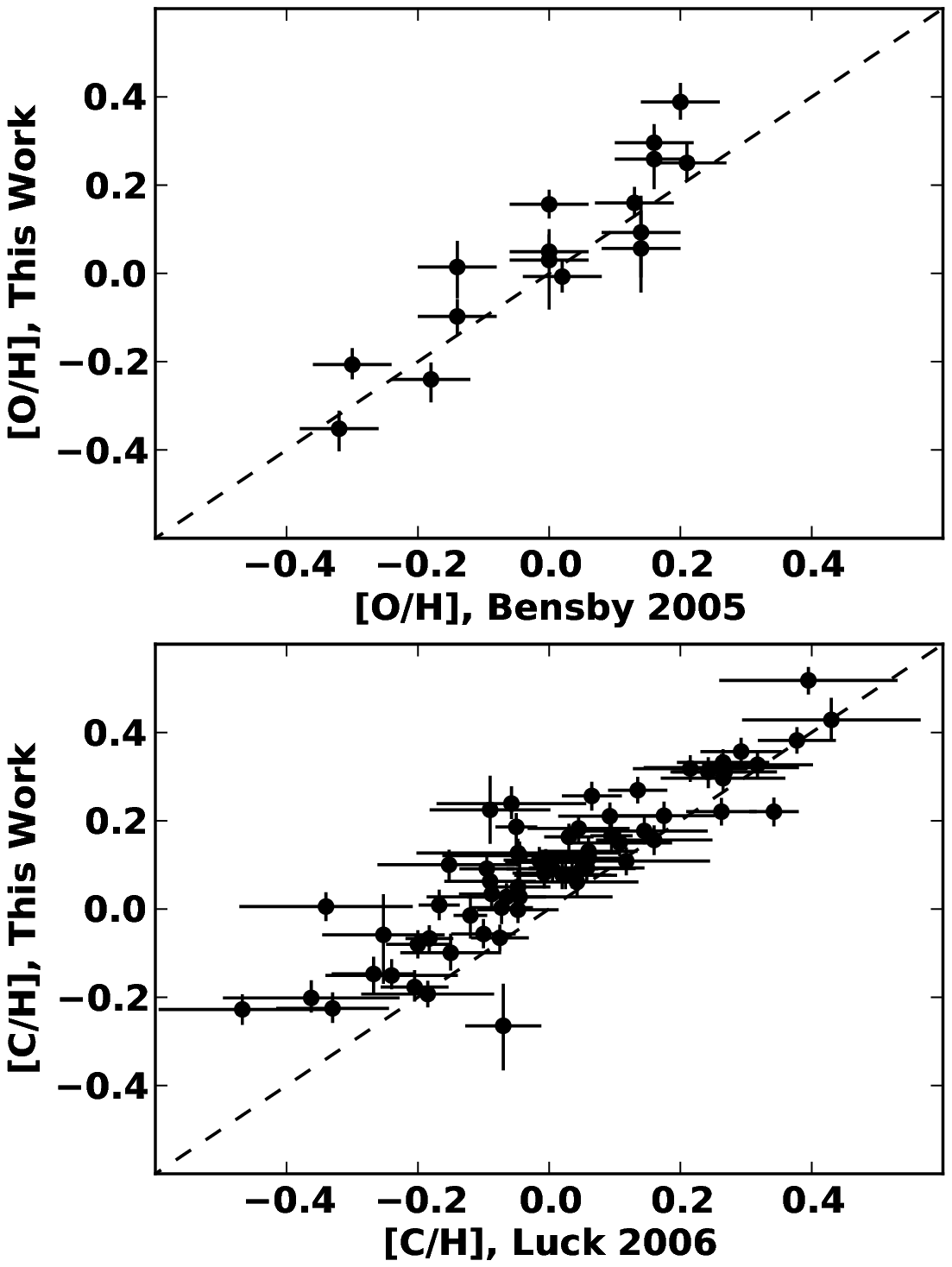}
\end{center}
\caption{Comparison plots of abundances from this work against those
  of~\cite{Bensby05} and~\cite{Luck06}.  The line shows a 1:1
  correlation.  We note a systematic offset in the carbon comparison.
  This may stem from the fact that the two works use different
  indicators.}
\label{fig:comp}
\end{figure}

We compare our results with~\cite{Bensby05} and~\cite{Luck06}.  We
report oxygen abundances for~\nCompO~stars analyzed by~\cite{Bensby05}
and \nCompC~stars analyzed by~\cite{Luck06}. We plot the comparison in
Figure~\ref{fig:comp}.  Our results track these comparison studies
well.  We recognize that the agreement is poorest for low values of
[C/H].  This likely the result of less robust fits to stars with
weaker carbon features.

The standard deviation of the differences in derived abundances is
\StdCompO~dex for oxygen and \StdCompC~dex for carbon.  Since
\cite{Bensby05} and \cite{Luck06} use different instruments,
spectral synthesizers, and fitting algorithms, it is unlikely there
are common systematic errors.  Therefore, the scatter in the
differences can be interpreted as a measure of the typical combined
statistical and systematic error.  We cannot say how much of the
observed scatter is due to our errors and those of the comparison
studies.

\subsection{Abundance Trends in the Thin and Thick Disks}
The Milky Way is thought to be made up of three distinct star
populations: The thin disk, thick disk, and the halo.  Most of the
stars in the local neighborhood belong to the thin disk, which has a
scale height of 300 pc.  The thick disk has a scale height of 1450 pc
and is comprised of older, metal-poor stars.  

\cite{PeekThesis} combined proper motion measurements from the {\em Hipparcos}
catalog~\citep{ESA97} with radial velocity measurements from the 
\cite{Nidever02}, SPOCS, and N2K catalogs into three-dimensional space motions
for \nThreeD~of our \nSampStars~program stars.  \cite{PeekThesis} computed
the probability of membership to each of the three populations in the manner
of \cite{Bensby03}, \cite{Mishenina04}, and \cite{Reddy06} for \nPop~of our 
\nStarsTot~stars with measured carbon and oxygen.
Our sample contains \nThin~thin disk stars, \nThick~thick disk stars, 
\nHalo~halo stars, and \nBdr~borderline stars (all three membership
probabilities less than 0.7).

\begin{figure}
\begin{center}
\includegraphics{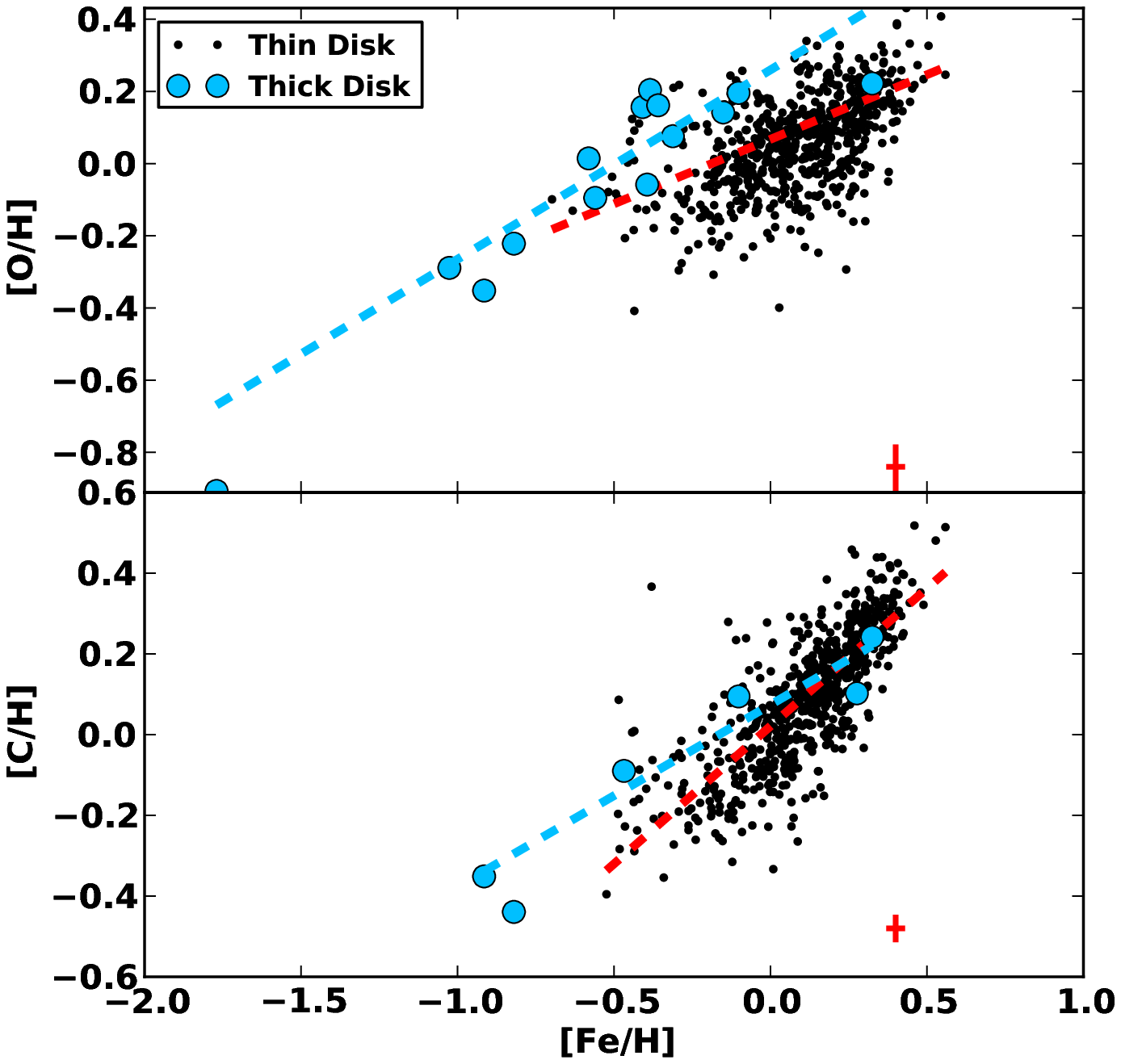}
\end{center}
\caption{ Carbon and oxygen abundance plotted against iron abundance.
  The \colorone\ points are the thin disk stars; the \colortwo\ points
  are the thin disk stars.  The line shows the abundance ratios in 0.1
  dex bins.  The crosses show median uncertainties.}
\label{fig:xonh}
\end{figure}

We plot [O/H] and [C/H] against [Fe/H] in Figure~\ref{fig:xonh}.  We
fit the trends with a line and list the best fit parameters in
Table~\ref{tab:xonh}. If the scatter was purely statistical, we would
expect our fits to have a reduced-$\sqrt{\chi^2}\sim1$.  Our fits have
reduced-$\sqrt{\chi^2}\sim2$, which suggests that some of the observed
scatter is astrophysical.  These main sequence stars have not begun to
process heavy elements, so the ranges of C, O, and Fe ratios reflect
the heterogeneous interstellar medium from which they formed.

\begin{deluxetable}{l c c c c c}
\tablewidth{0pc}
\tablecaption{Best fit parameters to abundance trends.}
\tablehead{
\colhead{}    &
\colhead{Pop.}&
\colhead{$m$} &
\colhead{$b$} &
\colhead{$\sqrt{\chi^2}$} \\
}
\startdata
\input{xonh.tex}
\enddata
\tablecomments{We fit thin and thick disk abundance trends with the
  following function [X/H] = $m$ [X/Fe] + $b$.  The best fit
  parameters are listed above along with the reduced-$\sqrt{\chi^2}$.}
\label{tab:xonh}
\end{deluxetable}

\begin{figure}
\begin{center}
\includegraphics{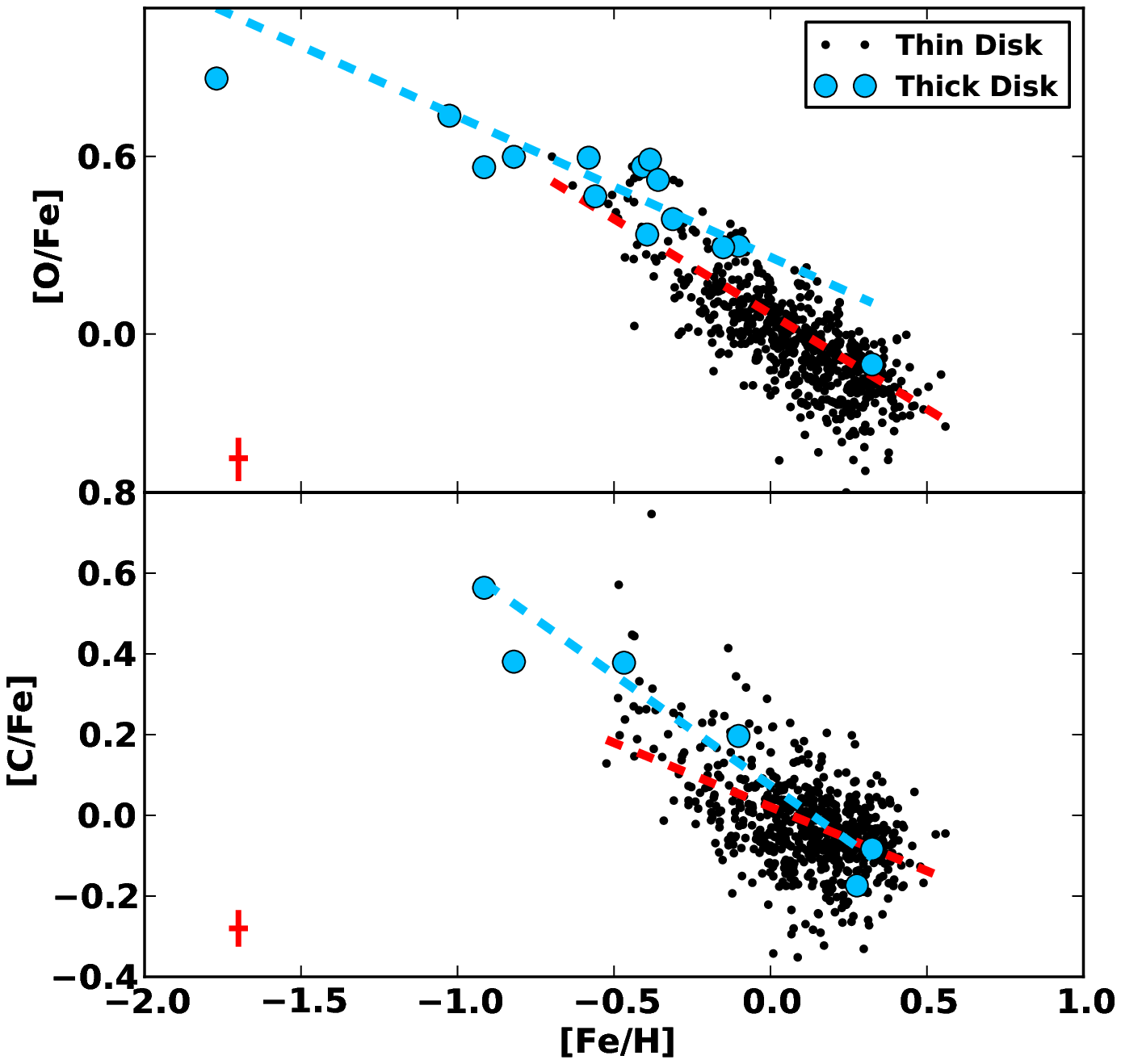}
\end{center}
\caption{ The ratios of carbon and oxygen to iron plotted against iron
  abundance.  The \colorone\ points are the thin disk stars; the
  \colortwo\ points are the thin disk stars.  The line shows the
  average ratios in 0.1 dex bins.  The crosses show median
  uncertainties.}
\label{fig:xonfe}
\end{figure}

We also plot [O/Fe] and [C/Fe] against [Fe/H] in
Figure~\ref{fig:xonfe}. The trends suggest that carbon and oxygen
lagged behind iron production for much of the period of galactic
chemical enrichment.  These trends flatten out for high [Fe/H].  Due
to the paucity of thick disk stars in our sample, we are cautious in
interpreting its abundance trends.  However, in the thick disk, oxygen
seems to be enhanced relative to iron, a result also reported by
\cite{Bensby04}.  This enhancement in oxygen suggests that type II
supernova played a more active role in enriching the thick disk.

\subsection{Exoplanets}
\hostFen~stars in our initial~\nSampStars~sample are known to host
planets.~\cite{Gonzalez97} measured relatively high stellar
metallically in the first four exoplanet host stars,
and \cite{Santos04} and \cite{Fischer05} showed that the fraction of stars bearing planets
increases rapidly above solar metallicity.  In light of the
correlation between C, O, and Fe, it is not surprising that hosts to
exoplanets are enriched in carbon and oxygen relative to the
comparison sample.

As shown in Table~\ref{tab:exo}, the mean [O/H] of the planet host and
comparison sample is~\hostOm~dex and~\compOm~dex respectively.  If we
take the error on the mean abundance to be the standard deviation of
derived abundances divided by the square root of the
number of stars in each sample i.e.
$\sigma_{mean} = \frac{\mathrm{Std. \ Dev.}}{\sqrt{N}}$, $\sigma_{mean}$
for [O/H] is 0.01 dex.  Carbon is also enriched in planet hosts 
where the mean [C/H]
is \hostCm~dex ($\sigma_{mean} = 0.02$ dex)
compared to \compCm~dex in the comparison sample with.
For both carbon and oxygen, the mean abundance
of the planet host sample is enriched by $\sim 5 \sigma $ compared
to the non-host sample.

\begin{figure}
\begin{center}
\includegraphics{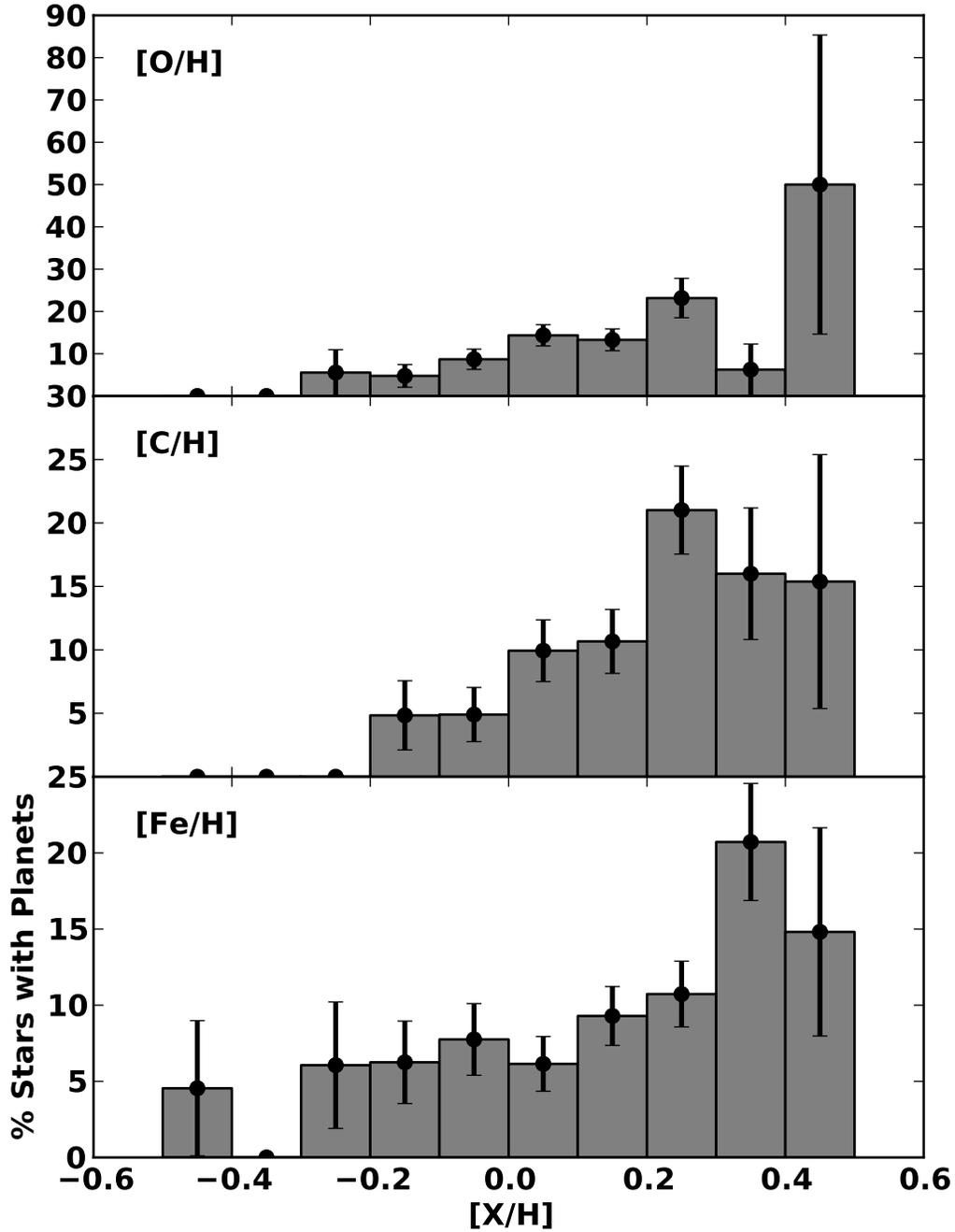}
\end{center}
\caption{The percentage of stars with known planets for 0.1 dex bins in oxygen, carbon, and iron.  The histograms are constructed from the \nStarsO, \nStarsC, and \nSampStars~stars with reliable measurements of [O/H], [C/H], and [Fe/H] respectively.}
\label{fig:exo}
\end{figure}

In Figure~\ref{fig:exo}, we divide the stars into 0.1 dex bins in
[X/H].  For each bin, we divide the number of planet-bearing stars by
the total number of stars in the bin.  As with iron, we observe an
increase in planet occurrence rate as carbon and oxygen abundance
increases.  While there is a hint of a possible plateau or turnover at
the highest abundance bins, these bins are dominated by small number
statistics. The data are not inconsistent with a monotonic rise,
within the errors.  The possibility that very enriched systems inhibit
planet formation is intriguing, and this parameter space warrants
further exploration.

\begin{deluxetable}{l  c c c c  c c c c c c}
\tablewidth{0pc}
\tablecaption{Statistical abundance properties of stars with planets. }
\tablehead{
\multicolumn{1}{c}{} &
\multicolumn{4}{c}{Hosts} &
\multicolumn{1}{c}{} &
\multicolumn{4}{c}{Non-Hosts} &
\multicolumn{1}{c}{}
 \\
 \cline{2-5} \cline{7-10}
&
\colhead{N} &
\colhead{mean} &
\colhead{Std. Dev.} &
\colhead{$\sigma_{mean}$} &
&
\colhead{N} &
\colhead{mean} &
\colhead{Std. Dev.} &
\colhead{$\sigma_{mean}$}
}
\startdata
\input{exo.tex}
\enddata
\label{tab:exo}
\tablecomments{We list the number of stars, mean abundance (dex), standard deviation (dex), and error on the mean abundance (dex) for the host and non-host populations.  The error on the mean abundance is computed by
$\sigma_{mean} = \frac{\mathrm{Std. \ Dev.}}{\sqrt{N}}$.
}
\end{deluxetable}

\subsection{$N_{C}/ N_{O}$}
We present the ratio of carbon to oxygen atoms, $N_{C}/ N_{O}$%
\footnote{$N_{C}/ N_{O} = 10^{\logepsc -\logepso}$}
for~\ncoStarsTot~stars with reliable
carbon and oxygen measurements as listed in the last column of
Table~\ref{tab:stellardata} of the Appendix.
Since we do not report carbon for stars cooler than~\teffCutClo~K,
our $N_{C}/ N_{O}$ measurements apply only to F and G spectral types.
While there is a weak
correlation between $N_{C}/ N_{O}$ and Fe at high [Fe/H], we note the large
degree of scatter in $N_{C}/ N_{O}$, which spans a wide range
from~\coMin~to~\coMax.

We emphasize that our measurements of [C/H] and [O/H] are differential
relative to solar and should be insensitive to revisions in the solar
abundance distribution.  $N_{C}/ N_{O}$ depends on our adopted solar
abundances of of oxygen \citep{Scott09} and carbon \citep{Caffau10}.  
We believe the abundances of carbon and oxygen are known at the $\sim 0.1$ dex
level.  Therefore, we expect revisions to the solar abundance distribution to 
systematically shift our $N_{C}/ N_{O}$ measurements by roughly $\sim 10^{0.1}$ or $\sim 25\%$ 

We measure~\ncoGtThresh~stars with $N_{C}/ N_{O}$ greater than~\coThresh.
Given the size of our random errors as determined by the Monte Carlo analysis of 
\S~\ref{sec:staterr}, very few of these stars are $1 \sigma$ detections of 
$N_{C}/ N_{O} > 1.$  However, since these errors are random, we believe our
measurements accurately reflect the {\em distribution} of $N_{C}/N_{O}$ in 
nearby disk stars.  Neglecting the zero-point offsets discussed earlier, we 
measure $N_{C}/ N_{O} > 1$ for roughly 10\% of nearby FG stars.

As noted by our anonymous reviewer, the CO molecule controls the equilibrium between carbon and oxygen in M dwarfs.  It is believed that $N_{C}/N_{O} > 1$ in M dwarfs results in an atmosphere rich in C$_{2}$, while $N_{C}/N_{O} < 1$ 
gives rise to TiO.  We are unaware of M dwarfs with strong C$_{2}$ bands indicating $N_{C}/N_{O} > 1$. This suggests such a
population is rare, assuming we understand the behavior of carbon-rich
M dwarf atmospheres.
We also note the additional complexities involved in modeling M star atmospheres.  Abundance estimates in cool stars rely on opacity tables of H$_{2}$O and 
other molecules that are not well understood at the temperatures
probed by M star atmospheres.
The fact that M stars are fully convective and have strong magnetic fields 
introduce additional complexities into model atmospheres.

Despite the uncertainties in accurately measuring $N_{C}/ N_{O}$, we have characterized the distribution of $N_{C}/ N_{O}$ for an unprecedented number of FG stars.  Furthermore, we have identified~\ncoGtThresh\ stars have high $N_{C}/ N_{O}$.
Given the predictions regarding exotic planets that form in a carbon rich
environment, these stars constitute important hosts for future work on
their exoplanets and exozodiacal dust.  Observations of dust with ALMA
and JWST may be particularly valuable.

\begin{figure}
\begin{center}
\includegraphics{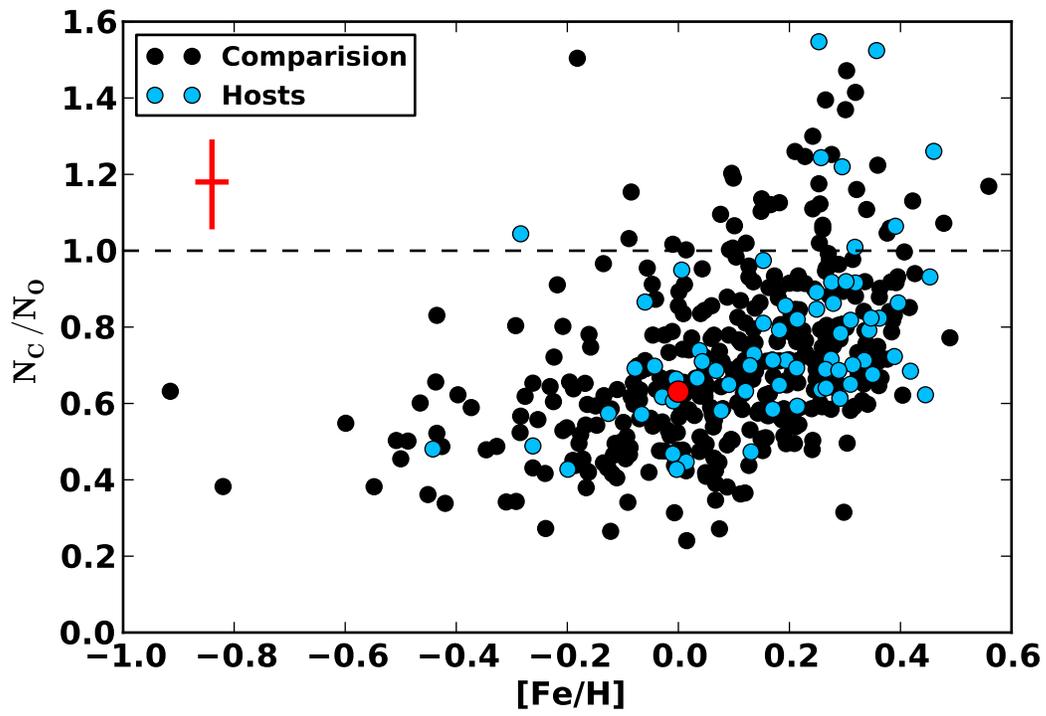}
\end{center}
\caption{$N_{C}/ N_{O}$ as a function of iron
  abundance.  The large \colorthree\ dot shows the solar values. The
  horizontal line shows equal carbon and oxygen.}
\label{fig:cofe}
\end{figure}

\subsection{WASP-12}
WASP-12b, discovered by~\cite{Hebb09}, is a transiting gas giant and a
favorable target for atmosphere studies.  \cite{Campo11} and
\cite{Croll11} measured secondary eclipses of WASP-12b at wavelengths
ranging from 1-8 \um, which can be used to characterize the planet's
dayside emission spectrum.  In a recent study, \cite{Madhu11} found
that these measurements are best-described by atmosphere models with
$N_{C}/ N_{O}$ $\geq$ 1 at 3 sigma significance.

We analyze WASP-12 identically to the~\nSampStars~star sample.  With a
V-mag of~\waspVmag~\citep{Hebb09}, WASP-12 is dimmest star in this
work and our spectra have S/N $\sim$ 50.  However, we measured oxygen
and carbon based on~\waspNobsO~and~\waspNobsC~spectra respectively.
We find [O/H] = \waspO, [C/H] = \waspC, and $N_{C}/ N_{O}$ = \waspCO\ i.e.,
sub-solar.

If the composition of the host star truly reflects the material from
which WASP-12b formed, our measurements suggest that WASP-12b does not
have a carbon-dominated bulk composition.  It is possible that the
planet acquired extra carbon at some point during its formation, or
that the planet's nightside and/or interior are acting as a sink for
oxygen, creating a carbon-rich dayside atmosphere while maintaing bulk
$N_{C}/ N_{O}$ less than unity.  In any case, this planet and its host star
warrant further study.

\section{Conclusion}
We have presented oxygen and carbon abundances for \nStarsTot~stars
based on HIRES spectra gathered by the Keck telescope.  We measure
oxygen by fitting the reliable 6300~\AA~forbidden line with {\tt SME}
and self-consistently account for the significant nickel blend.  Our
carbon abundances are derived from the 6587~\AA~ CI line.  Our errors
are based on a rigorous Monte Carlo treatment, and our measurements
agree with values in the literature.  Our sample is large enough to
characterize and remove systematic trends due to \teff.  We see that
carbon and oxygen are both enriched in stars with known planets.  We
see a significant number of stars with $N_{C}/ N_{O}$ exceeding unity, which
supports the possibility that some stars host exotic carbon-rich
planets.  However, our measurement of sub-solar $N_{C}/ N_{O}$ for WASP-12,
complicates the recent claim by \cite{Madhu11} that WASP-12b is a
carbon world.

\acknowledgements The authors are indebted to Kathryn M. G. Peek,
Andrew McWilliam, Debra A. Fischer, Heather Knutson, and Martin Asplund
for productive and enlightening conversations that improved this work.  
We thank the many observers who collected the Keck
HIRES data used here: Debra A. Fischer, Jason T. Wright, John Asher
Johnson, Andrew W. Howard, Chris McCarthy, Suneet Upadhyay, R. Paul
Butler, Steven S. Vogt, Eugenio Rivera, Joshua Winn, Kathryn
M. G. Peek, and Howard Isaacson. We gratefully acknowledge the
dedication of the staff at Keck Observatory, particularly Grant Hill
and Scott Dahm for their HIRES support.  We acknowledge salary support
for Petigura from the Thye family through the Berkeley Summer
Undergraduate Research Fellowship program and by the National Science
Foundation through the Graduate Research Fellowship Program.  This
research has made use of the SIMBAD database, operated at CDS,
Strasbourg, France; the Vienna Atomic Line Database; the Kurucz Atomic
and Molecular Line Databases; the NIST Atomic Spectra Database; the
Exoplanet Orbit Database; the Exoplanet Data Explorer at
exoplanets.org, and NASA's Astrophysics Data System Bibliographic
Services. The authors extend thanks to those of Hawaiian ancestry on
whose sacred mountain of Mauna Kea we are privileged to be
guests. Without their generous hospitality, the Keck observations
presented here would not have been possible.
\bibliographystyle{apj}
\bibliography{manuscript}

\appendix
\input{stellardata.tex}
\end{document}

%% file: texcmd.tex
\newcommand{\nThin}{847} 
\newcommand{\nThick}{16} 
\newcommand{\nHalo}{12} 
\newcommand{\nBdr}{25} 
\newcommand{\nPop}{900} 
\newcommand{\vsiniCutO}{7} 
\newcommand{\maxTO}{0.11} 
\newcommand{\nStarsO}{694} 
\newcommand{\vsiniCutC}{15} 
\newcommand{\teffCutClo}{5300} 
\newcommand{\maxTC}{0.15} 
\newcommand{\nStarsC}{704} 
\newcommand{\coThresh}{1.00} 
\newcommand{\scatterCut}{0.30} 
\newcommand{\teffSol}{5770} 
\newcommand{\StdCompC}{0.09} 
\newcommand{\StdCompO}{0.08} 
\newcommand{\nCompC}{67}   
\newcommand{\nCompO}{16}   
\newcommand{\compCm}{0.08} 
\newcommand{\compOm}{0.05} 
\newcommand{\hostCm}{0.17} 
\newcommand{\hostFen}{100} 
\newcommand{\hostOm}{0.10} 
\newcommand{\nStarsTot}{941} %
\newcommand{\ncoGtThresh}{46} %
\newcommand{\coMax}{1.55} %
\newcommand{\ncoStarsTot}{457} %
\newcommand{\coMin}{0.24} %

%% file: abundhist.tex
$ {[}O/H] $& 694 & 0.06 & 0.14 & -0.91 & 0.43\\$ {[}C/H] $& 704 & 0.09 & 0.17 & -0.52 & 0.52\\$ {[}Fe/H] $& 1070 & 0.07 & 0.27 & -1.95 & 0.56\\

%% file: xonh.tex
[C/H] & thick & 0.450 $\pm$ 0.074 & 0.074 $\pm$ 0.035 & 2.19 \\

[C/H] & thin & 0.682 $\pm$ 0.019 & 0.021 $\pm$ 0.004 & 2.52 \\

[O/H] & thick & 0.525 $\pm$ 0.081 & 0.260 $\pm$ 0.047 & 2.41 \\

[O/H] & thin & 0.358 $\pm$ 0.017 & 0.067 $\pm$ 0.004 & 1.66 \\

%% file: exo.tex
{[}O/H] & 88 & 0.10 & 0.12 & 0.01 & 
         & 606 & 0.05 & 0.14 & 0.01 & \\

{[}C/H] & 79 & 0.17 & 0.14 & 0.02 & 
         & 625 & 0.08 & 0.17 & 0.01 & \\

{[}Fe/H] & 100 & 0.17 & 0.18 & 0.02 & 
         & 970 & 0.06 & 0.27 & 0.01 & \\

%% file: stellardata.tex
\begin{deluxetable}{rrrrrrrrrrrrrrrrr}
\tabletypesize{\scriptsize}
\setlength{\tabcolsep}{0.05in}

\tablecaption{Stellar Data.}
\tablewidth{0pt}
\tablehead{
  \\[1ex]
  \colhead{Name\tablenotemark{a}}
  & \colhead{$V$\tablenotemark{b}}
  & \colhead{$d$\tablenotemark{b}}
  & \colhead{T \tablenotemark{c}}
  & \colhead{$\log{g}$\tablenotemark{c}}
  & \multirow{2}{*}{Pop \tablenotemark{d}}	
  & \multirow{2}{*}{Host}
  & \multirow{2}{*}{[M/H] \tablenotemark{c,e}}
  & \multirow{2}{*}{[Ni/H] \tablenotemark{c}}	
  & \multicolumn{1}{c}{\multirow{2}{*}{$N_{O}$}}
  & \multicolumn{1}{c}{\multirow{2}{*}{[O/H]\tablenotemark{f}} }
  & \multicolumn{1}{c}{\multirow{2}{*}{$N_{C}$}}
  & \multicolumn{1}{c}{\multirow{2}{*}{[C/H]\tablenotemark{f}} }
  & \multicolumn{1}{c}{\multirow{2}{*}{C/O\tablenotemark{f}} }
  \\
  \colhead{(HD)}
  & \colhead{(mag)}
  & \colhead{(pc)}
  & \colhead{(K)}
  & \colhead{(cgs)}
  &
  &
  & 
  & 
  &
  &
  &
  &
  \\
}
\startdata
\input{bigtable.tex}
\enddata
\tablenotetext{a}{Names are HD numbers unless otherwise indicated.}
\tablenotetext{b}{Johnson $V$-magnitude and distance from the Hip
   catalog; where stars were not in the Hip
   catalog, $d$ and $V$ are from the Geneva-Copenhagen Catalog 
   \citep{Nordstrom04}; $d$ and $V$ are taken from SIMBAD 
   as a last resort.}
\tablenotetext{c}{Stellar parameters from VF05 and D.~Fischer.}
\tablenotetext{d}{Membership identification taken from \cite{PeekThesis}}
\tablenotetext{e}{[M/H] is computed from all elements heavier than He.}
\tablenotetext{f}{We have corrected these abundances a for a systematic trend in temperature (see \S~\ref{sec:abundances}).}

\label{tab:stellardata}
\end{deluxetable}
%


%% file: bigtable.tex
\\[0ex] 105 & 7.51 & 40 & 6126 & 4.65 & dn & no &-0.02 & 0.02 & 0 & \nd & 4 & $-0.04_{-0.04}^{+0.04}$ & \nd \\ 
 \\[0ex] 166 & 6.07 & 13 & 5577 & 4.58 & dn & no &0.12 & 0.19 & 11 & $-0.09_{-0.08}^{+0.07}$ & 7 & $0.12_{-0.04}^{+0.03}$ & $1.02_{-0.19}^{+0.21}$ \\ 
 \\[0ex] 377 & 7.59 & 39 & 5873 & 4.28 & dn & no &0.11 & 0.12 & 0 & \nd & 10 & $-0.09_{-0.03}^{+0.03}$ & \nd \\ 
 \\[0ex] 449 & 8.58 & 95 & 5860 & 4.40 & dn & no &0.12 & 0.27 & 1 & $0.23_{-0.07}^{+0.07}$ & 3 & $0.38_{-0.03}^{+0.03}$ & $0.90_{-0.15}^{+0.16}$ \\